Superconductivity in an Einstein Solid $A_xV_2Al_{20}$ (A = Al and Ga)

Atsushi ONOSAKA, Yoshihiko OKAMOTO, Jun-ichi YAMAURA, and Zenji HIROI[*]

*Institute for Solid State Physics, University of Tokyo, 5-1-5 Kashiwanoha, Kashiwa, Chiba 277-8581, Japan*

[*]E-mail address: hiroi@issp.u-tokyo.ac.jp

A cage compound $A_xV_2Al_{20}$ ($Al_{10}V$), that was called an Einstein solid by Caplin and coworkers 40 years ago, is revisited to investigate the low-energy, local vibrations of the A atoms and their influence on the electronic and superconducting properties of the compound. Polycrystalline samples with A = Al, Ga, Y, and La are studied through resistivity and heat capacity measurements. Weak-coupling BCS superconductivity is observed below $T_c$ = 1.49, 1.66, and 0.69 K for $A_x$ = $Al_{0.3}$, $Ga_{0.2}$, and Y, respectively, but not above 0.4 K for $A_x$ = La. Low-energy modes are detected only for A = Al and Ga, which are approximately described by the Einstein model with Einstein temperatures of 24 and 8 K, respectively. A weak but significant coupling between the low-energy modes, which are almost identical to those called rattling in recent study, and conduction electrons manifests itself as anomalous enhancement in resistivity at around low temperatures corresponding to the Einstein temperatures.

About 40 years ago, Caplin and coworkers called the intermetallic compound $Al_{10}V$ an Einstein solid,[1,2] because it exhibits a unique local mode associated with excess Al atoms that are loosely bound to their neighbor and can move around inside the surrounding cage with low frequencies. This local vibration was characterized by an Einstein mode with a sharply defined Einstein temperature $T_E$ of 23 ± 1 K. They also found a lower-energy mode with $T_E$ = 10 K for smaller Ga atoms placed inside the cage in Ga-doped $Al_{10}V$.[2] Legg and Lanchester found an anomalously large lattice expansion in the same compounds, which was ascribed to the volume dependence of local mode excitations.[3] These vibrations are probably identical to those now called rattling after recent extensive study on three cage compounds; Si-Ge clathrates,[4] filled-skutterudites,[5] and β-pyrochlore oxides.[6-8]

Rattling is a local and essentially anharmonic oscillation with a large atomic excursion of an atom confined in an oversized atomic cage in crystals.[8] Since structural coupling between a guest atom (rattler) and the surrounding cage is too weak to generate a dispersive mode expected for an ordinary crystal, rattling can be a local mode to be approximated by an Einstein mode within the harmonic approximation. Note, however, that there must be serious deviation from an Einstein mode particularly for an extreme case with large anharmonicity such as observed in $KOs_2O_6$;[7,8] this deviation may be a central topic to be investigated in the research of rattling. As size mismatch between rattler and cage increases, the first excited level goes down to result in an unusually low-energy excitation, which can be approximately measured by Einstein temperature $T_E$. Compared with $T_E$s thus far obtained for the three cage compounds ($k_B T_E$ = 2-7 meV), those for $Al_{10}V$ are very low, especially when Ga atoms are introduced. Thus, one expects more intense rattling in the present compounds.

What is fascinating about rattling is an electron-rattler (e-r) interaction that causes interesting electronic properties. It has been clearly demonstrated in previous study on β-pyrochlore oxides that a large e-r interaction gives rise to a peculiar, concave-downward temperature dependence in resistivity at high temperatures and $T^2$ resistivity at low temperatures.[6] Moreover, it causes ratting-mediated superconductivity, particularly an extremely strong-coupling superconductivity with $T_c$ = 9.60 K in $KOs_2O_6$.[7] It is considered that the e-r interaction is enhanced by the anharmonicity of rattling.[8]

We have noticed, in the course of our study on β-pyrochlore oxides, a pioneering work by Caplin *et al*. on $Al_{10}V$ and found it interesting to revisit the compound in the light of recent understanding of rattling phenomena. The compound was reported by Carlson *et al*. in 1955 as the most Al-rich phase in the V-Al binary system and called phase α or $VAl_{11}$.[9,10] The crystal structure was determined to be cubic with lattice constant $a$ = 1.45~1.46 nm in the space group $Fd$-$3m$.[11,12] As two groups independently decided the occupancy $g$ at the midst of the cage as zero ($Al_{10}V$) or 0.5 ($Al_{10.25}V$), it has been believed that there is a range of solid solutions between them. Later, Caplin and Nicholson estimated the number of oscillators inside the cage by heat capacity measurements and obtained $g$ = 0.3 ~ 0.7.[2] Thus, the cage is only partially occupied by excess Al atoms owing to some reason, although the range of occupancy is still unknown. In this study, we call this cage compound $A_xV_2Al_{20}$, instead of $Al_{10}V$ or others, to explicitly express the number of atoms inside the cage as $x$ on the basis of the $CeCr_2Al_{20}$ structure: $x$ = 0 and 1 mean zero and full occupations of the cage by the A atom, respectively.

Provided atomic positions in the space group $Fd$-$3m$ with origin choice 2, as assumed in previous studies on the family of $CeCr_2Al_{20}$-type compounds,[13] the structure of $A_xV_2Al_{20}$ has V atoms in the 16$d$ positions and Al atoms in the 96$g$, 48$f$, 16$c$ and 8$a$ positions, as shown in Fig. 1.

The 8$a$ position with the $T_d$ site symmetry is located at the center of the largest cage made of four 16$c$ Al and twelve 96$g$ Al atoms. The Al atom in the 8$a$ position has unusually long bonds of 0.31-0.32 nm to the neighbors forming the cage, compared with 0.28 nm for the interatomic distance in metallic aluminum. Caplin *et al*. found that this 8$a$ Al atom can move around inside the cage and behaves as an Einstein oscillator with a sharply defined characteristic temperature of 23 K.[1,2] However, their resistivity measurements revealed that the scattering due to this mode is less pronounced so that coupling to conduction electrons is rather weak. Claeson and Ivarsson reported superconductivity at 1.6-1.7 K for compounds containing either Al or Ga atoms inside the cage without giving any experimental data.[14] They mentioned that there is little evidence to show the contribution of the low-energy modes to the superconductivity.

In the present study, we aimed to investigate the superconductivity in more detail and to reveal whether or not the low-energy modes affect the electronic properties of the compounds. We synthesized a series of polycrystalline samples of $A_xV_2Al_{20}$ with A = Ga, Al, Y, and La, the metallic radius of the A atoms being increased from Ga to La, and examined them through heat capacity and resistivity measurements. We show that there is a little but significant effect of low-energy modes or rattling on the electronic properties of the A = Al and Ga compounds.

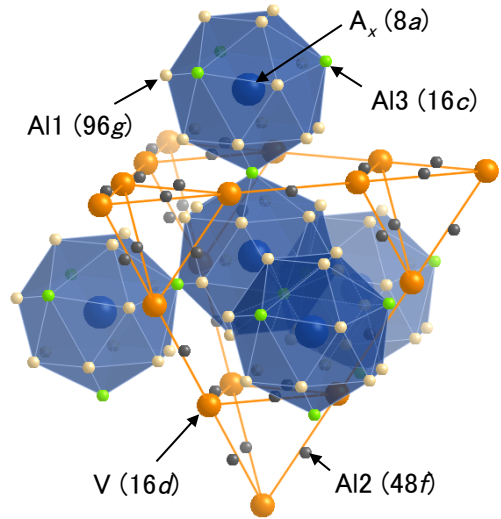

Fig. 1. (Color online) Crystal structure of $A_xV_2Al_{20}$ in the $CeCr_2Al_{20}$ structure with the space group $Fd$-3$m$ (origin choice 2). A rattling A atom at the 8$a$ position is confined in a cage made of four 16$c$ Al and twelve 96$g$ Al atoms. The sets of 8$a$ Al and V atoms form diamond and pyrochlore lattices, respectively.

Polycrystalline samples of $A_xV_2Al_{20}$ with A = Al were prepared by a solid-state reaction from aluminum and vanadium in nominal ratios of $x$ = 0-1; preparation of a single crystal is difficult, because the compound decomposes incongruently above 670 ºC.[10] First, a composite was melted at high temperature in an arc-melt furnace to obtain a uniform mixture. After sealed in an evacuated quartz ampoule, the mixture was annealed at 650 ºC for 80 h for the compound to form by a peritectic reaction. Powder X-ray diffraction measurements indicate that monophasic samples are obtained only near $x$ = 0.3 in this condition: samples with smaller and larger $x$ values contained $Al_{45}V_7$ and Al as impurity phases, respectively; the two impurity phases are superconductors with $T_c$ = 0.9 and 1.2 K, respectively.[14] This suggests that solubility range is narrower than reported before.[11,12] We examined the physical properties of the monophasic sample with $x$ = 0.3 for A = Al. Samples with A = Ga, Y and La were prepared in similar methods by annealing at 640, 700, and 800 ºC, respectively. Samples with A = Ga seemed to be monophasic in a wide range of $x$, but here we examined only the $x$ = 0.2 sample; Ga composition dependences of superconducting and other properties will be reported elsewhere. For A = Y and La, stoichiometric samples with $x$ = 1 were prepared; deficiency at the 8$a$ site must be negligible. The lattice constants of the cubic unit cell are 1.45157(8), 1.45171(5), 1.45386(8), and 1.46222(9) nm for $A_x$ = $Al_{0.3}$, $Ga_{0.2}$, Y, and La, respectively. The small and systematic increase in lattice constant is consistent with the corresponding increase in the metallic radius of A atoms in the common rigid cage. The chemical compositions of the A = $Al_{0.3}$ and $Ga_{0.2}$ samples were examined by the energy-dispersive X-ray analysis in a scanning electron microscope, which confirmed that the intended compositions were nearly retained in the products. Resistivity and heat capacity were measured in a PPMS (Quantum Design).

Figure 2 shows superconducting transitions observed in resistivity and heat capacity. Sharp drops in resistivity are observed at zero-resistivity temperatures of 1.50, 1.68 and 0.70 K for $A_x$ = $Al_{0.3}$, $Ga_{0.2}$ and Y, respectively. Correspondingly, second-order transitions are clearly detected in heat capacity at nearly the same temperatures. The mean-field $T_c$s, which are defined as a midpoint of a jump in heat capacity by taking into account the entropy balance, are 1.49, 1.66 and 0.69 K, respectively. The superconductivity of $YV_2Al_{20}$ may be reported here for the first time, as far as we know. No superconductivity is observed above 0.4 K for A = La. Superconducting and other parameters obtained from heat capacity measurements are listed in Table I. Note, in Fig. 2, that the magnitudes of heat capacity just above $T_c$ for A = Al and Ga are markedly enhanced compared with those for A = Y and La, indicating large contributions from low-energy phonons even at this low temperature region.

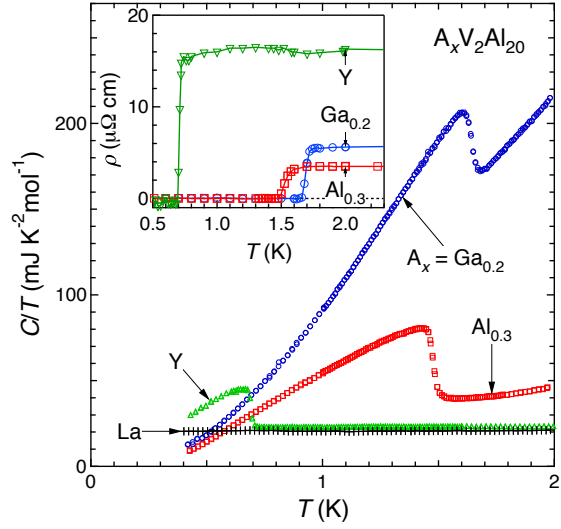

Fig. 2. (Color online) Heat capacity divided by temperature $C/T$ and resistivity of polycrystalline samples of $A_xV_2Al_{20}$ ($A_x$ = $Al_{0.3}$, $Ga_{0.2}$, Y, and La) showing superconducting transitions at $T_c$ = 1.49, 1.66 and 0.69 K for A = $Al_{0.3}$, $Ga_{0.2}$, and Y, respectively, and no transition above 0.4 K for A = La.

Superconductivity was completely suppressed above 0.4 K in a magnetic field of 0.3 T: the upper critical fields of the A = Al and Ga samples are approximately 0.3 T at $T$ = 0. Electronic heat capacity $C_e$ and the Sommerfeld coefficient $\gamma$ are determined by subtracting data measured at 0.3 T from zero-field data. Figure 3 compares $C_e/T$ normalized by $\gamma$ for the three superconducting compounds. The curves nearly overlap to each other, showing a typical form expected for a weak-coupling BCS superconductivity. The magnitudes of the jump at $T_c$ are 1.35-1.41 (Table I), close to a theoretical value of 1.43. Moreover, the data of A = Al and Ga at $T/T_c < 0.5$ can be well fitted to an exponential form of $\exp(-\Delta/k_BT)$, yielding $2\Delta/k_BT_c$ = 2.74 and 2.90, respectively; these values are slightly smaller than a theoretical value of 3.53. Thus, $s$-wave superconductivity occurs in these compounds, as is generally expected for a phonon-based superconductivity.

The observed increase in $T_c$ from 0.69 K (Y) to 1.49 K (Al) and further to 1.66 K (Ga) may be related to the corresponding decrease in the energy of low-energy phonons or the increase in their contributions at low energy, as observed in heat capacity shown in Fig. 2. $\gamma$ also becomes large accordingly, as listed in Table I. Thus, the low-energy phonons must play an important role in the superconductivity of $A_xV_2Al_{20}$, although a systematic study as a function of the A content is required for further discussion. Nevertheless, the superconductivity remains in the weak coupling regime as observed in $CsOs_2O_6$ with the weakest rattling in the family of β-pyrochlore oxides, and a strong-coupling superconductivity such as observed in the strongest-rattling compound $KOs_2O_6$ is not realized in $A_xV_2Al_{20}$.

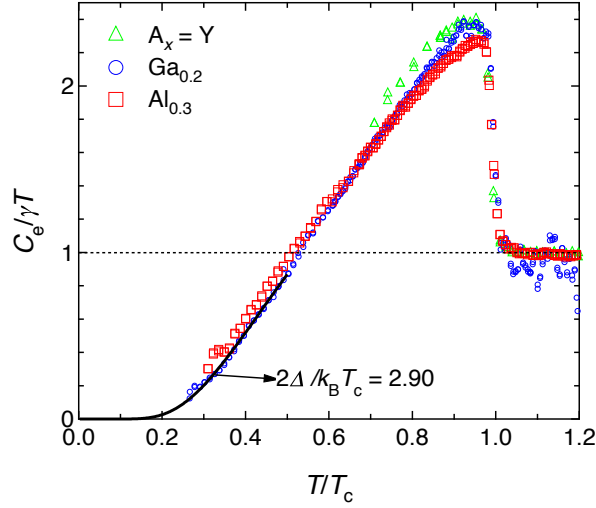

Fig. 3. (Color online) Electronic heat capacity divided by temperature normalized by the Sommerfeld coefficient $C_e/\gamma T$ as a function of normalized temperature $T/T_c$ for $A_xV_2Al_{20}$ ($A_x$ = $Al_{0.3}$, $Ga_{0.2}$, and Y). The values of $T_c$ and $\gamma$ are listed in Table I. The thick line below $T/T_c = 0.5$ shows a fit to an exponential form for the $Ga_{0.2}$ data, yielding $2\Delta/k_BT_c = 2.90$.

Next we investigate lattice contributions in heat capacity. Figure 4(a) shows heat capacity divided by temperature below 300 K. Obviously, there are large enhancements below 30 K for $Al_{0.3}$ and $Ga_{0.2}$ than for Y and La (Y data is not shown in the figure). The $C/T$ of the Y and La compounds are proportional to $T^2$ below ~10 K, as shown in the inset of Fig. 4(a) for La in a $C/T$ vs. $T^2$ plot, indicating that the lattice heat capacity is well described by the Debye model with reasonable Debye temperatures of $T_D = 420$ and 430 K, respectively; the vibrations of Y and La atoms take large energies and may be incorporated into normal lattice vibrations. The presence of additional contributions for the Ga and Al compounds indicates that Ga and excess Al atoms present inside the cage vibrate with low energies. A common contribution to heat capacity from the $V_2Al_{20}$ framework can be estimated from the $C/T$ of $LaV_2Al_{20}$ after subtraction of $\gamma$ and multiplication by a factor of 22/23. The contribution from A-atom vibrations $C(A)/T$ shows a broad peak at ~10 K for Al, as shown in Fig. 4(b), which is well reproduced by assuming a single Einstein mode with a number of oscillators $\delta = 0.280(3)$ per formula unit and an Einstein temperature $T_E = 23.7(2)$ K. The estimated number of oscillators is nearly equal to the nominal composition $x$, which means that only 30% cages are occupied by Al atoms. The $T_E$ is similar to those reported by Caplin et al.[2] and Legg et al.[3], and the $\delta$ equals the smallest value that Caplin et al. reported.[2] As mentioned earlier, Al$_x$V$_2$Al$_{20}$ seems to exist in a narrow composition range near $x = 0.3$ in our preparation condition.

On the other hand, the $C(A)/T$ of the Ga$_{0.2}$ sample exhibits a complex peak shape that is expanded to lower temperatures compared with that of Al$_{0.3}$. We can fit the data by assuming two Einstein modes, *i.e.* ($\delta$, $T_E$) = (0.050(2), 8.1(1) K) and (0.250(3), 23.4(3) K). The former is presumably due to smaller Ga atoms contained in the cage and the latter due to Al atoms, because the latter $T_E$ is nearly equal to that in the Al pure sample. The $T_E$ of a Ga atom is also in good agreement with those previously reported.[1-3] However, the presence of both Ga and Al atoms in the cages of a Ga-doped compound has not been noticed in the previous studies. Our analyses clearly reveal that 0.05 Ga atoms per formula unit are present together with 0.25 Al atoms in cages, and the rest of Ga atoms, *i.e.* 0.15, should occupy the cage position. The total number of atoms inside the cage seems to be ~0.3, similar to the case of the pure Al sample. There must be a specific reason for this magic number, possibly related to band filling. The smaller $T_E$ of Ga, which is heavier than Al, is in line with what is expected for a harmonic oscillator. In contrast, in the case for β-pyrochlore oxides, $T_E$ becomes smaller from Cs to K with decreasing the atomic weight.[6] This unusual tendency has been ascribed to decreasing ionic radius in a rigid cage and thus increasing the guest-free space as well as anharmonicity. We think that the very small $T_E$ of Ga is also due to a similar effect, as Ga is smaller than Al in metal radius; 0.135 nm for Ga and 0.143 nm for Al.

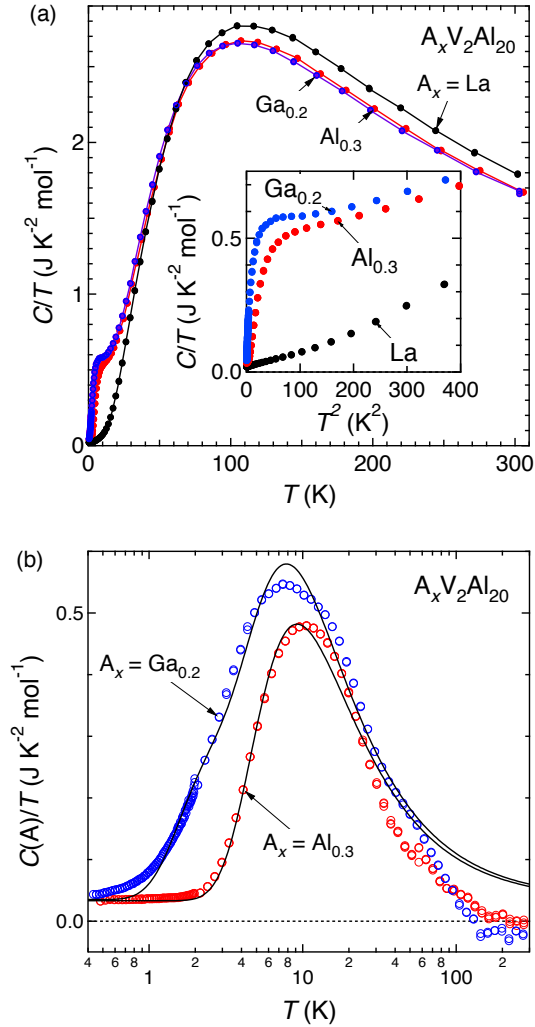

Fig. 4. (Color online) (a) Heat capacity divided by temperature for $A_xV_2Al_{20}$ ($A_x$ = $Al_{0.3}$, $Ga_{0.2}$, and La) in a wide temperature range. The low-temperature part is expanded in the inset and plotted as a function of $T^2$. (b) Heat capacity of A-atom vibrations inside the cage $C(A)$ after subtraction of the cage contribution estimated from the data of $LaV_2Al_{20}$. The data between 0.4 and 300 K are shown. Curves on the $Al_{0.3}$ and $Ga_{0.2}$ data in (b) represent fits assuming one and two Einstein modes, respectively.

Interestingly noted in Fig. 4(b) is the presence of a deviation from the Einstein behavior at low temperature observed only for $Ga_{0.2}V_2Al_{20}$: the $C(A)/T$ is obviously enhanced from the fitting curve below 1.5 K. The temperature dependence below 1.5 K is nearly proportional to $T^2$, suggesting a Debye-like heat capacity with a small Debye temperature of 34 K. Note that it has remained after the subtraction of the contribution from the normal Debye modes ($T_D$ = 430 K). This additional contribution must come from unusual low-energy excitations that are related to the vibrations of Ga atoms inside the cage but are different from a simple Einstein mode. This deviation may reflect a large anharmonicity of Ga atoms inside the cage and will be studied in the future.

A possible interaction between the above-mentioned A-atom vibrations and conduction electrons should manifest itself in the temperature dependence of resistivity at low temperatures, because such low-energy modes can give rise to large scattering of carriers even at low temperatures where few normal phonons are generated. The resistivities of the $Al_{0.3}$ and $Ga_{0.2}$ compounds are shown in Fig. 5, together with that of $YV_2Al_{20}$. The Y compound shows a saturating behavior in $\rho$ as $T$ approaches zero, which is approximately proportional to $T^3$, possibly indicating of two contributions from a weak electron-electron (proportional to $T^2$) and a normal electron-phonon scatterings (proportional to $T^5$).[15] In contrast, the Al and Ga compounds exhibit almost $T$-linear behavior below ~ 30 K at first glance. However, a careful examination makes it clear that there exist broad humps in the resistivity at approximately 20 and 10 K for the Al and Ga compounds, respectively. Note that these temperatures are close to the corresponding Einstein temperatures, suggesting that scatterings by the low-energy rattling modes are responsible for the enhancements. Therefore, electron-rattler interactions in $A_xV_2Al_{20}$ are weak but in fact present, serving as scatterers even at such low temperatures.

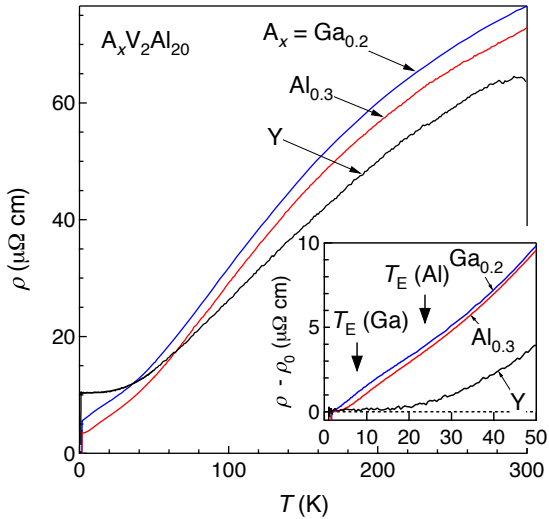

Fig. 5. (Color online) Temperature dependences of resistivity for polycrystalline samples of $A_xV_2Al_{20}$ ($A_x$ = $Al_{0.3}$, $Ga_{0.2}$, and Y). The inset shows $\rho - \rho_0$, where $\rho_0$ is the resistivity just above $T_c$.

Caplin *et al*. suggested the possibility of off-center rattling for $A_xV_2Al_{20}$, which is in contrast to the on-center rattling observed in β-pyrochlore oxides:[6-8] the on-center position in the cage of $A_xV_2Al_{20}$ has a higher potential energy than surrounding off-center positions, so that the guest atom can move around along the surface of the cage and behave much more as a "rotator".[2] We consider that such off-center rattling vibrations are responsible for the observed enhancement in resistivity. Relevance of them to the Debye-like tail in heat capacity observed only for the Ga compound is interesting to be studied in the future.

In summary, we have studied cage compound $A_xV_2Al_{20}$ to understand the rattling of the A atoms and its effects on the electronic and superconducting properties. Weak-coupling BCS superconductivity is observed below $T_c$ = 1.49, 1.66 and 0.69 K for $A_x$ = $Al_{0.3}$, $Ga_{0.2}$ and Y, respectively, but not above 0.4 K for A = La. Low-lying Einstein-like modes are detected only for A = Al and Ga with Einstein temperatures of 24 and 8 K, respectively. A weak but significant coupling between the rattling modes and conduction electrons manifests itself as unusual enhancements in resistivity at around low temperatures corresponding to the Einstein temperatures.

**Acknowledgments**

We are grateful to T. Kobayashi, K. Ishida, and H. Harima for helpful discussions. This work was supported by Grant-in-Aids for Scientific Research on Priority Areas "Heavy Electrons" (No. 23102704) provided by MEXT, Japan.

Table I. Superconducting, electronic, and lattice properties of $A_xV_2Al_{20}$: $A_x$ = $Al_{0.3}$, $Ga_{0.2}$, Y, and La.

| | $Al_{0.3}V_2Al_{20}$ | $Ga_{0.2}V_2Al_{20}$ | $YV_2Al_{20}$ | $LaV_2Al_{20}$ |
|---|---|---|---|---|
| $T_c$ (K) | 1.49 | 1.66 | 0.69 | < 0.4 |
| $\Delta C/\gamma T_c$ | 1.35 | 1.40 | 1.41 | - |
| $2\Delta(0)/k_B T_c$ | 2.74 | 2.90 | - | - |
| $\gamma$ (mJ K$^{-2}$ mol$^{-1}$) | 33 | 35 | 27 | 20 |
| $T_E$ (K) | 23.7(2) | 8.1(1) / 23.4(3) | 420* | 430* |
| $\delta$ | 0.280(3) | 0.050(2) / 0.250(3) | - | - |

*Debye temperature for $YV_2Al_{20}$ and $LaV_2Al_{20}$.